\documentclass[prl,twocolumn,showpacs,nofootinbib,floats,floatfix,superscriptaddress]{revtex4}

\usepackage{graphicx}
\usepackage{amsmath,amsfonts}
\usepackage{dcolumn}
\usepackage{color}

\newcommand{\td}[1]{{\rm d}#1} % total derivative
\newcommand{\SMetric}{\gamma} % spatial 3-metric
\newcommand{\CF}{\psi} % conformal factor
\newcommand{\ExCurv}{K} % spatial extrinsic curvature
\newcommand{\BMetric}{h} % boundary metric
\newcommand{\BCD}{D} % boundary covariant derivative
\newcommand{\BRicciS}{{{}^2\!R}} % boundary Ricci scalar
\begin{document}

\title{Approximate Killing Vectors on $S^2$}

\author{Gregory B. Cook}\email{cookgb@wfu.edu}
\affiliation{Department of Physics, Wake Forest University,
		 Winston-Salem, North Carolina\ \ 27006}
\author{Bernard F. Whiting}\email{bernard@phys.ufl.edu}
\affiliation{Department of Physics, University of Florida, 
                 Gainesville, FL \ \ 32611}

\date{\today}

\begin{abstract}
We present a new method for computing the best approximation to a
Killing vector on closed 2-surfaces that are topologically $S^2$.
When solutions of Killing's equation do not exist, this method is
shown to yield results superior to those produced by existing methods.
In addition, this method appears to provide a new tool for studying
the horizon geometry of distorted black holes.
\end{abstract}


\maketitle

\section{Introduction}
\label{sec:introduction}

An exact geometric sphere possesses a two parameter family of
rotational Killing vectors while even a slightly distorted sphere may
possess no Killing vectors whatsoever.  Nevertheless, one could
imagine defining perturbations of these  initial vectors which
would, in some well-defined ``best'' sense, represent the closest
available approximation to vectors which almost satisfy Killing's
equation on such a slightly distorted sphere.  In this paper we
introduce a definition which is best in a least squared sense and
discuss some of its attributes.

In general relativity, rotational Killing vectors play an important
role in providing a quasi-local definition for the spin of a rotating
body.  A system of astrophysical interest, such as a pair of orbiting
black holes, possesses no global rotational Killing vectors.  In this
case, angular momentum can only be rigorously defined for the system
as a whole in terms of asymptotic rotational Killing vectors.  

A quantity of great importance in the evolution of black-hole binaries
is the spin of the individual black holes.  The spin of such black
holes can only be determined by some approximate quasi-local
definition (see Ref.\cite{Szabados-2004} for a review).  There exist
many different quasi-local definitions for angular momentum, but they
all take the form of an integral over a 2-surface with topology $S^2$,
and all require a rotational Killing vector on this surface.

In numerical relativity, the quasi-local spin of a black hole is most
often expressed as
\begin{equation}
\label{eq:QL_spin0}
S_{(\xi)}=\frac{1}{8\pi G}\oint_{\cal S}
  \ExCurv_{ij}\xi^j\,d^2S^i,
\end{equation}
where $\ExCurv_{ij}$ is the extrinsic curvature of a spatial
hypersurface with metric $\SMetric_{ij}$, $d^2S^i$ is the area element
of an $S^2$ surface of integration ${\cal S}$ taken to be a black
hole's apparent horizon, and $\xi^i$ is a Killing vector of the metric
$\BMetric_{ij}$ induced on ${\cal S}$ by $\SMetric_{ij}$.  This form
was derived by Brown and York\cite{brown-york-1993} and later within
the Isolated Horizons framework (see Ref.\cite{Ashtekar-Krishnan-2004}
for a review).  It gives the angular momentum of the rotation
associated with the rotational Killing vector $\xi^i$.  Unfortunately,
the induced metric $\BMetric_{ij}$ will not admit a solution of
Killing's equation for the case of orbiting black hole binaries.  In
this situation, one has no recourse but to find some reasonable
approximation for the Killing vector required in
Eq.~(\ref{eq:QL_spin0}).

In some cases, conformal Killing vectors have been used, and have
yielded physically reasonable results\cite{caudill-etal-2006}.  Better
still, a ``Killing Transport'' (KT) technique\cite{dreyer-etal-2003},
which finds exact Killing vectors when they are present, has been
recently adopted, and appears to give physically reasonable results
(cf Refs.~\cite{caudill-etal-2006,Campanelli-etal-2007a}).  Our
definition is shown to be even better in a well-defined least squared
sense and, for coalescing binary black holes, yields other interesting
results worthy of further investigation.

\section{Equations For the Best Approximate Killing Vector}
\label{sec:AKV}

An arbitrary vector field $\xi^i$ on $S^2$ can be decomposed into
two scalars $d$ and $v$
\begin{equation}
\label{eq:gen_xi_dcmp}
  \xi^i \equiv \BCD^i d + \epsilon^{ij}\BCD_j\upsilon,
\end{equation}
where $\BCD_i$ is the covariant derivative compatible with the metric
$\BMetric_{ij}$ induced on the surface ${\cal S}$ and $\epsilon_{ij}$
is the Levi-Civita tensor.  Similarly, the general gradient of a
1-form can be expressed as
\begin{equation}
\label{eq:gen_dxi_dcmp}
  \BCD_i\xi_j \equiv L\epsilon_{ij} + \BMetric_{ij}\Lambda + S_{ij},
\end{equation}
where $L$ and $\Lambda$ are scalars, and $S_{ij}$ is symmetric and
trace-free.  Eqs.~(\ref{eq:gen_xi_dcmp}) and (\ref{eq:gen_dxi_dcmp})
immediately imply that
\begin{eqnarray}
\label{eq:Lambda_Def_1}
  \Lambda &=& \mbox{$\frac12$}\BCD^i\BCD_i d, \\
\label{eq:L_Def_1}
  L &=& -\mbox{$\frac12$}\BCD^i\BCD_i v, \\
\label{eq:S_Def_1}
  S_{ij} &=& \BCD_i\BCD_j d 
            - \mbox{$\frac12$}\BMetric_{ij}\BCD^k\BCD_k d \\
	    && \mbox{} \nonumber
	    + \mbox{$\frac12$}\left(
	       \epsilon_{ik}\BCD_j\BCD^k v 
	       + \epsilon_{jk}\BCD_i\BCD^k v\right).
\end{eqnarray}  

For $\xi^i$ to be Killing, it must satisfy Killing's equation
$\BCD_{(i}\xi_{j)}=0$ where the parentheses denote symmetrization.
This implies that $\Lambda$ and $S_{ij}$ must vanish if $\xi^i$ is
Killing.  We may choose $\Lambda$ to vanish, in which case
Eq.~(\ref{eq:Lambda_Def_1}) implies that $d$ is harmonic.  
But, assuming a non-singular metric, the
only harmonic function on $S^2$ is a constant, which makes no contribution
to $\xi^i$.  Thus, we are left with
\begin{eqnarray}
\label{eq:approx_xi_dcmp}
  \xi^i &\equiv& \epsilon^{ij}\BCD_j\upsilon, \\
\label{eq:approx_dxi_dcmp}
  \BCD_i\xi_j &\equiv& L\epsilon_{ij} + S_{ij}.
\end{eqnarray}

Our goal is to find an approximate Killing vector that minimizes the
non-Killing aspects of $\xi^i$.  Clearly, having already set $\Lambda$
to zero, a solution with $S_{ij}$ as close to zero as possible is what
we desire, so we proceed by finding a vector $\xi^i$ that minimizes
$S_{ij}S^{ij}$.

From Eqs.~(\ref{eq:approx_xi_dcmp}) and (\ref{eq:approx_dxi_dcmp})
it follows that
\begin{equation}
\label{eq:Ldxi-ident}
  L = \mbox{$\frac12$}\epsilon_{ij}\BCD^i\xi^j,
\end{equation}
and
\begin{equation}
\label{eq:SS_def_1}
  S_{ij}S^{ij} = (\BCD_i\BCD_j v)(\BCD^i\BCD^j v) 
   - \mbox{$\frac12$}(\BCD^k\BCD_k v)^2.
\end{equation}
Now, we wish to extremize $S_{ij}S^{ij}$ with respect to $v$.  But, we
must do this in a way that is independent of the {\rm normalization}
of $\xi^i$.  Using $|\xi|^2 = (\BCD_iv)(\BCD^iv)$, we choose the following
scalar function on ${\cal S}$
\begin{equation}
\label{eq:Lagrangian}
  {\cal L} \equiv S_{ij}S^{ij} 
  + \mbox{$\frac12$}\BRicciS\Theta(\BCD_k v)(\BCD^k v),
\end{equation}
where $\BRicciS$ is the Ricci scalar associated with $\BMetric_{ij}$
and $\Theta$ is a dimensionless constant.  Varying with respect to
$v$, $\delta{\cal L}/\delta{v}=0$ yields a fourth-order scalar elliptic
equation for $v$ that can be rewritten as a pair of second-order
scalar elliptic equations for $v$ and $L$
\begin{eqnarray}
\label{eq:L_appKe_gen}
  \BCD^i\BCD_i L 
   - (1-\Theta)\left[\mbox{$\frac12$}(\BCD^i\BRicciS)\BCD_i v 
     - \BRicciS L\right] &=& 0, \\
\label{eq:v_appKe_gen}
  \BCD^i\BCD_i v + 2L &=& 0.
\end{eqnarray}
Eqs.~(\ref{eq:L_appKe_gen}) and (\ref{eq:v_appKe_gen}) can be solved
for $L$ and $v$, with the Lagrange multiplier $\Theta$ fixed by the
requirement that Eq.~(\ref{eq:L_appKe_gen}) be integrable on $S^2$.
Given a solution for $v$, the approximate rotational Killing vector is
given by Eq.~(\ref{eq:approx_xi_dcmp}) once it is normalized to have
affine length $2\pi$.  We note that Eq.~(\ref{eq:L_appKe_gen}) is
satisfied by a Killing vector $\xi^i$ when $\Theta=0$ (with $L$
defined by Eq.~(\ref{eq:Ldxi-ident}) and
$\BCD_iv=(2/\BRicciS)\BCD_iL$).

It has been noted\cite{Schnetter-etal-2006,Hayward-2006a} that an
approximate Killing vector should be divergence free $\BCD_i\xi^i=0$,
in part because the angular momenta computed using such an approximate
Killing vector possess a certain gauge invariance.  Approximate Killing
vectors constructed using the KT method are not guaranteed to be
divergenceless, although at least in certain
cases\cite{caudill-etal-2006} this can be enforced {\em a posteriori}.
These approximate Killing vectors also inherit an additional problem.
A defining equation of the KT method\cite{dreyer-etal-2003}, which can
be written in the form of Eq.~(\ref{eq:Ldxi-ident}), is not satisfied
by solutions of the KT equations unless there is a Killing vector.
This is due to the path dependence of the solution scheme.  Moreover,
Eq.~(\ref{eq:Ldxi-ident}) {\em cannot} be enforced {\em a posteriori}.

Because our approximate Killing vector is defined from the solution of
Eqs.~(\ref{eq:L_appKe_gen}) and (\ref{eq:v_appKe_gen}) via
Eq.~(\ref{eq:approx_xi_dcmp}), it is guaranteed to be divergenceless.
Furthermore, because our solution is obtained via a global solution of
elliptic equations, Eq.~(\ref{eq:Ldxi-ident}) is also guaranteed to be
satisfied.

\section{Tests}
\label{sec:tests}

We have implemented a code to solve Eqs.~(\ref{eq:L_appKe_gen}) and
(\ref{eq:v_appKe_gen}) for situations where the metric $\BMetric_{ij}$
is conformal to a unit 2-sphere:
\begin{equation}
\label{eq:conf_metric}
\td{s}^2 = \CF^4 r^2(\td{\theta}^2 + \sin^2\theta\td{\phi}^2).
\end{equation}
Details of the solution scheme will be presented in a future
paper\cite{Cook-Whiting-2007b}.  While this form for the metric may
seem to be a strong simplification, a scheme based on
this form is suitably general since any sufficiently smooth metric on
$S^2$ is conformally equivalent to a unit 2-sphere.

To explore our method, we first consider the case where the conformal
factor is given by an $\ell=2$, $m=0$ scalar spherical harmonic with
its axis of symmetry rotated to a direction given by
$(\theta^\prime,\phi^\prime)$
\begin{equation}
\label{eq:psil2m0}
  \CF(\theta,\phi) = A + B\sum_{m=-2}^2{Y_{2m}(\theta,\phi)
                                        Y^*_{2m}(\theta^\prime,\phi^\prime)},
\end{equation}
and $A$ and $B$ are real constants chosen to guarantee that $\CF>0$
everywhere.  The form of Eq.~(\ref{eq:psil2m0}) guarantees that the
metric possesses a rotational Killing vector and in all the cases we
attempted, solving Eqs.~(\ref{eq:L_appKe_gen}) and
(\ref{eq:v_appKe_gen}) returned a solution where the axis of symmetry
was correctly rotated by $(\theta^\prime,\phi^\prime)$ and for which
$\Theta=0$.  As mentioned above, the solutions are divergenceless and
satisfy Eq.~(\ref{eq:Ldxi-ident}) to the level of roundoff error.
Furthermore, we find $S_{ij}S^{ij}=0$ also to the level of roundoff
error, as expected for a solution that yields a true Killing
vector.

Interestingly, Eqs.~(\ref{eq:L_appKe_gen}) and (\ref{eq:v_appKe_gen})
in general have {\em multiple solutions}.  In fact for the example
given by Eq.~(\ref{eq:psil2m0}), there exist an infinitely degenerate
set of solutions where the axis of approximate symmetry lies anywhere
in the rotated equatorial plane.  For these solutions, $\Theta\ne0$
and $S_{ij}S^{ij}\ne0$ since these solutions are not true Killing
vectors.  Again, the solutions are divergenceless and satisfy
Eq.~(\ref{eq:Ldxi-ident}) to the level of roundoff error.

\begin{figure}
\includegraphics[width=\linewidth,clip]{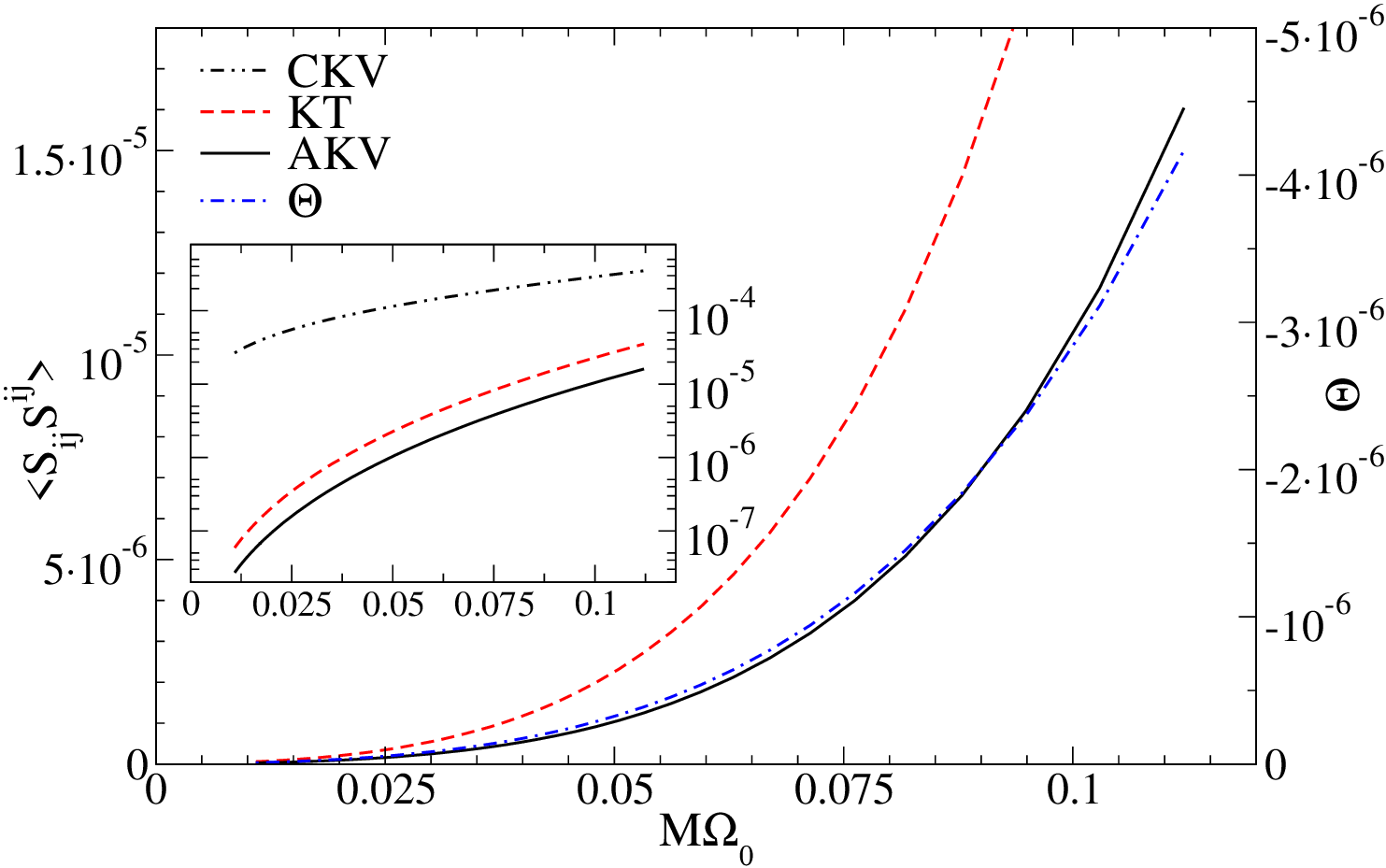}
\caption{\label{fig:CoSSTheta} The value of
$\langle{S_{ij}S^{ij}}\rangle$ for the new method is displayed as a
solid(black) line and its value when using the KT method is shown as a
dashed(red) line.  The analogous quantity $\langle{2\Lambda^2}\rangle$
when using conformal Killing vectors is shown as a
dot-dot-dashed(black) line visible in the inset where a logarithmic
scaling is used.  The value of the Lagrange multiplier $\Theta$
obtained using the new method is also displayed as a dot-dashed(blue)
line and its scale is shown on the right.  All data is from
one corotating black hole in an equal-mass binary.}
\end{figure}

We have also tested the system of equations against the numerically
generated initial data for corotating and non-spinning equal-mass
black-hole binaries as described in Ref.\cite{caudill-etal-2006}.  We
use data for different orbital separations, parameterized by the
dimensionless orbital angular velocity $M\Omega_0$, where $M$ is the
total irreducible mass of the binary.  For the case of corotating
black holes, the spin of each black hole is aligned with the direction
of the orbital angular momentum.  In all cases tested, we have found
that the measured spins based on an approximate Killing vector
obtained using Eqs.~(\ref{eq:L_appKe_gen}) and (\ref{eq:v_appKe_gen})
are {\em nearly identical} to those based on the KT method, with
differences growing only to a few parts in $10^7$.  While the
resulting spins are nearly identical, our solutions are measurably
different from the results of the KT method, and we find that our
system of equations produces solutions for which
$\langle{S_{ij}S^{ij}}\rangle \equiv
(4\pi)^{-1}\oint{S_{ij}S^{ij}d\Omega}$ is always smaller than that
produced by the KT method.  Note that in both cases, the approximate
Killing vectors have been properly normalized to have affine length
$2\pi$.  Figure~\ref{fig:CoSSTheta} shows
$\langle{S_{ij}S^{ij}}\rangle$ as computed by our method and the KT
method, and the analogous quantity $\langle{2\Lambda^2}\rangle$
obtained by using conformal Killing vectors.  On a separate scale,
Fig.~\ref{fig:CoSSTheta} also shows the value of $\Theta$ obtained by
our method for the same data.

\begin{figure}
\includegraphics[width=\linewidth,clip]{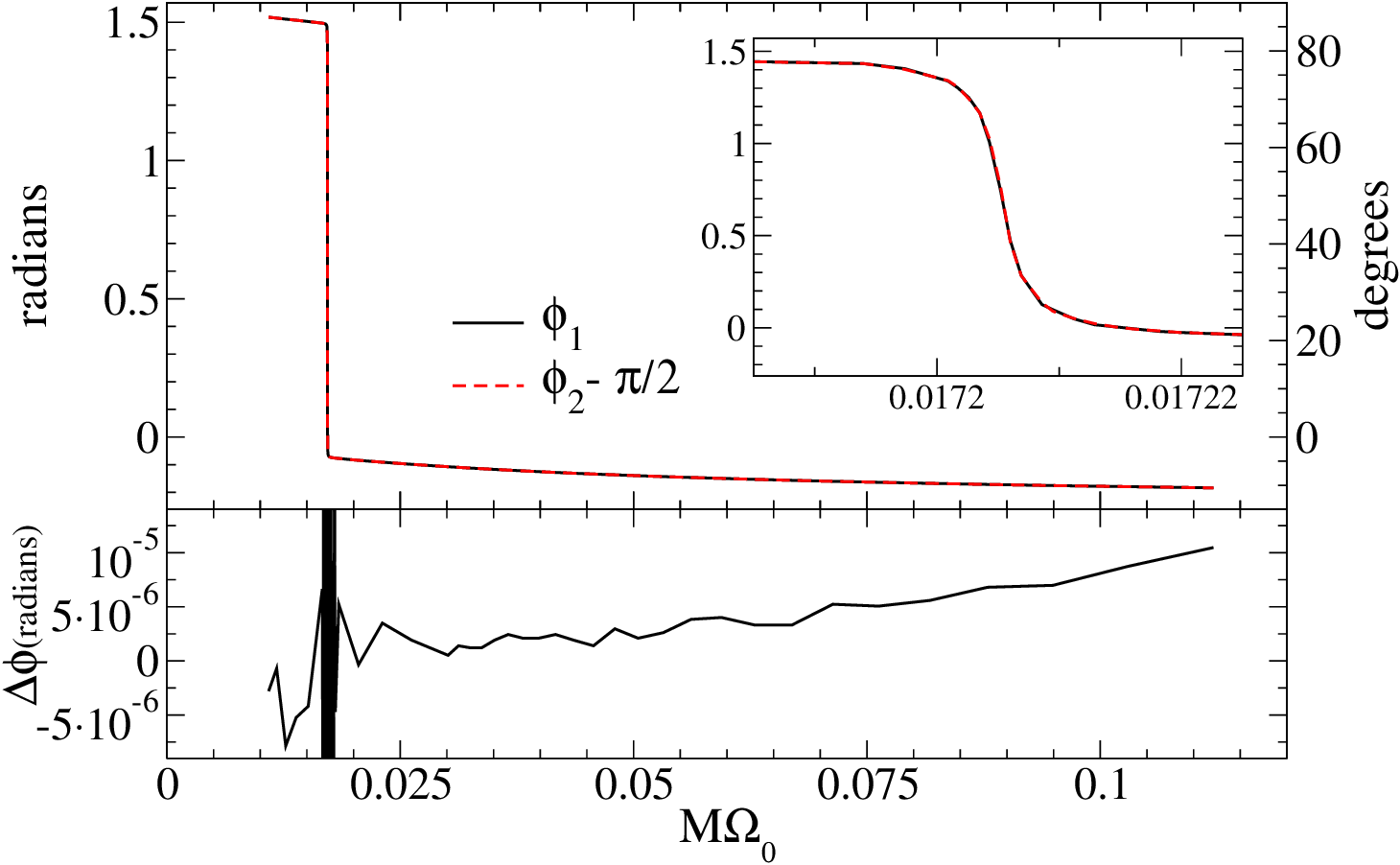}
\caption{\label{fig:Codphi} The top portion of the figure displays the
directions of the additional approximate symmetry axes with $\phi=0$
in the positive $x$ direction.  The axis direction of the Solution~1
is displayed as a solid(black) curve and the direction for Solution~2,
with $\pi/2$ subtracted, is displayed as a dashed(red) line.  The
lines are visually coincident.  The lower portion of the figure
displays $\Delta\phi\equiv\phi_1 -\phi_2 + \pi/2$ showing the degree
to which the two axes are approximately orthogonal.  All data is
from one corotating black hole in an equal-mass binary.}
\end{figure}

As with the first example given by Eq.~(\ref{eq:psil2m0}), our
equations yield multiple solutions for the numerically generated
black-hole data.  The solutions shown in Fig.~\ref{fig:CoSSTheta} have
their axis of approximate symmetry aligned with the orbital angular
momentum ($z$-axis) as expected.  In all cases, we find {\em two
additional solutions}.  Both have their axis of approximate symmetry
in the orbital plane and we refer to them as Solutions~1 and 2.  We
find that for all orbital separations, the directions of these
two symmetry axes are orthogonal to the level of numerical error in
the code (truncation error).  For large separation (small
$M\Omega_0$), Solution 1 has its axis of approximate symmetry pointed
roughly in the direction of motion of the black hole ($y$-axis) at
an azimuthal angle of $\phi_1$, and Solution 2 has its axis pointed
roughly toward the companion black hole ($x$-axis) at an angle of
$\phi_2$.  The directions $\phi_1$ and $\phi_2$ (mod $\pi$) for these
two solutions are displayed in Fig.~\ref{fig:Codphi}.  Here, zero
azimuthal angle is in the direction of the positive $x$-axis.  For
small separations, the roles are reversed and we find that Solution 1
points roughly toward the $x$-axis and Solution 2 toward the $y$ axis.

The regime between large and small orbital separations, where the two
solutions swap orientations, is quite interesting.  Over most of the
range of $M\Omega_0$ considered, $\phi_1$ and $\phi_2$ change
gradually.  However, over a narrow range of $M\Omega_0$, the angles
change rapidly but smoothly, rotating by an angle of approximately
$\pi/2$.  Interestingly, the value of $\langle{S_{ij}S^{ij}}\rangle$
for Solution 1 is smaller than that for Solution 2 for all
separations.  At the point where $\phi_1=\phi_2-\pi/2=\pi/4$, curves
of $\langle{S_{ij}S^{ij}}\rangle$ for the two solutions ``appear'' to
cross.  However, a careful examination shows this to be an ``avoided''
crossing.  We shall examine this behavior in more detail in a future
paper\cite{Cook-Whiting-2007b}.

Finally, if we measure the spin of the black holes using the
approximate Killing vectors associate with Solutions~1 and 2, all
cases yield zero to truncation error.  So, for the case of corotating
equal-mass black hole binaries, we find three ``orthogonal'' solutions
and only one of them yields a non-vanishing spin.

\begin{figure}
\includegraphics[width=\linewidth,clip]{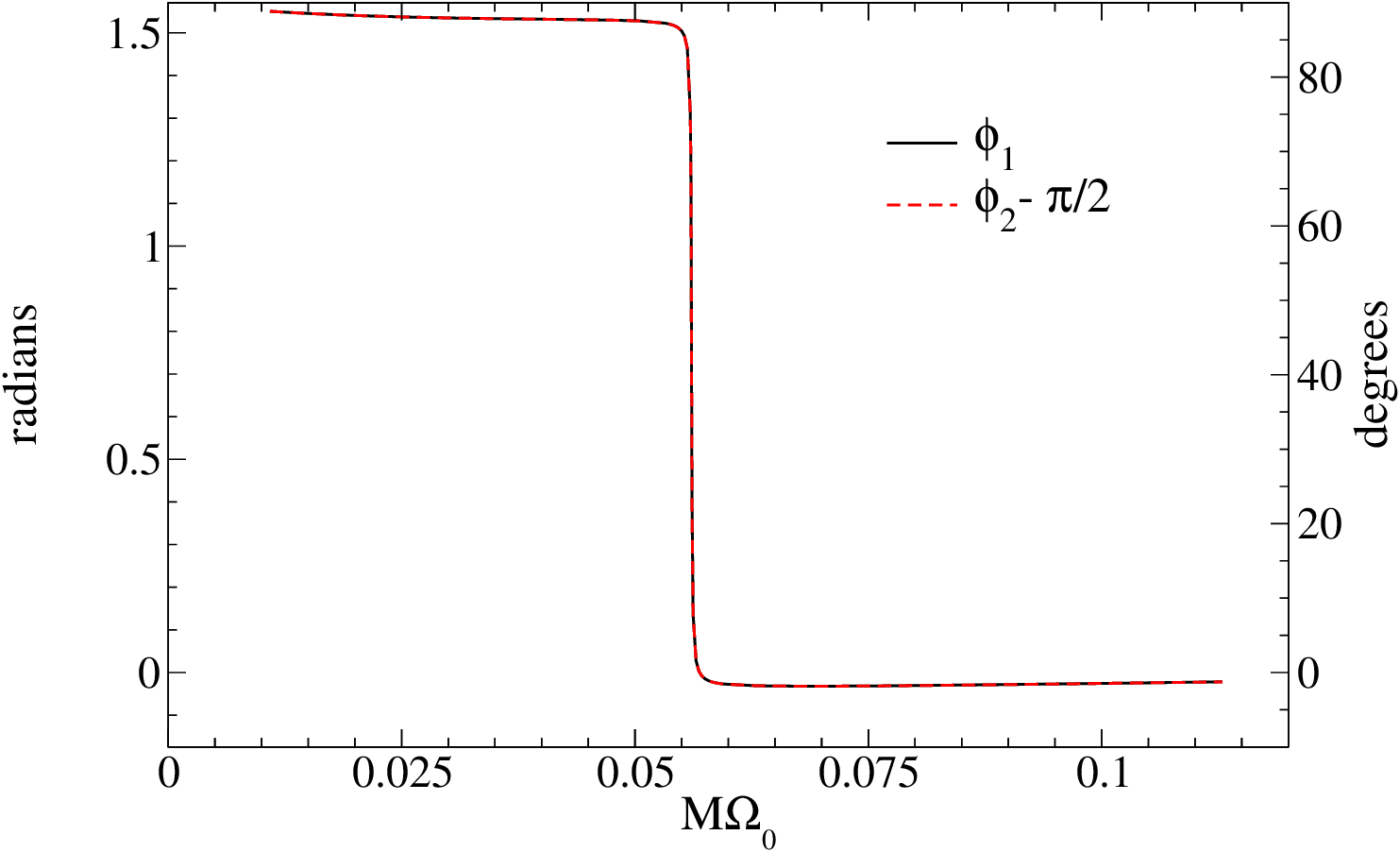}
\caption{\label{fig:Tidphi} The directions of the additional
approximate symmetry axes from one non-spinning black hole in an
equal-mass binary.  See description for Fig.~\ref{fig:Codphi}.}
\end{figure}

We find similar results for the case of non-spinning black-hole-binary
initial data.  There is a solution with an axis of approximate
symmetry aligned with the orbital angular momentum and two additional
solutions with their axes in the orbital plane and similarly
orthogonal to each other.  As discussed in
Ref.~\cite{caudill-etal-2006}, the non-spinning black hole data we use
are defined by setting the spin measured by the KT method to zero.  The
corresponding spins measured by our method again differ at most by a
few parts in $10^7$, and in all cases we find the value of
$\langle{S_{ij}S^{ij}}\rangle$ for the new method to be smaller than
the value from the KT solution.  The behavior of the additional
solutions is qualitatively the same as seen for the case of corotation
and is displayed in Fig.~\ref{fig:Tidphi}.  While the behaviors are
generally similar, the directions of the approximate symmetry axes are
somewhat different and the rapid change in the direction of the
solutions occurs at much smaller separation.

\section{Discussion}

We have mentioned two previous methods defined in the literature for
use in computing the spin of rotating black holes that lack axial
symmetry.  Both have been considered useful in the past, and yet both
have shortcomings.  When a Killing vector does not exist, the
conformal Killing approach returns a vector for which
$\BCD_{i}\xi^{i}\ne 0$.  By contrast, the Killing Transport method
constructs a $\xi^{i}$ that can often be made divergenceless.  It also
constructs the scalar $L$ (see Eq.~(\ref{eq:gen_dxi_dcmp})), but
generally this does not satisfy Eq.~(\ref{eq:Ldxi-ident}).  Our new
method not only ensures both that $\xi^{i}$ is divergenceless and that
Eq.~(\ref{eq:Ldxi-ident}) is satisfied, but it also is best in the
sense that $\langle{S_{ij}S^{ij}}\rangle$ is minimal.  Since a Killing
vector cannot be produced where one does not exist, the usefulness of
our results for an approximate Killing vector will depend on the
extent to which physical questions (such as concern black hole spins)
can be given meaningful answers.  In particular, it may have immediate
application in giving a more refined definition for binaries
containing black holes without individual spin.

One sense in which our results already appear meaningful relates to
the vectors found to reside in our $xy$-plane, for which the
corresponding spins are computed to be zero to the level of truncation
error.  Any other outcome for these would have been somewhat
unpalatable.  Their actual orientation for binary black holes at large
separation can be easily interpreted in terms of boosted frames.
Closer in, their combined dramatic rotation at some critical
separation warrants further investigation, as does the apparent
occurrence of avoided crossings for $\langle{S_{ij}S^{ij}}\rangle$ and
for $\Theta$ at the critical separation.

Another sense in which our results appear meaningful relates to the
solution associated with our $z$-direction.  For sufficiently large
black hole separation, our result leads to spins which are in good
agreement with those of the KT method.  This is not unreasonable, even
though we find small differences between the two solutions for $\xi^i$
over the surface of the apparent horizon.  However, for highly
distorted black holes --- such as near merger or with near maximal
rotation, or for unaligned spins --- we can imagine that our results
could prove to be more robust.  More extensive comparisons than we
have been able to carry out here will be necessary before this
expectation might be practically substantiated.

To a high degree of precision, it appears that the axes of the
approximate symmetries we find form an orthogonal basis in all the
cases we have examined.  This is of considerable interest, because
this basis is directly related to intrinsic properties of the apparent
horizons we have studied.  Thus, in addition to the reasonableness of
our results associated with both the rotation axis and the orbital
plane, our method appears to give a new tool for studying the horizon
geometry of distorted black holes in general, and the volatile
dynamics of black holes during collision in particular.  Further
investigation of the usefulness of this new tool will be forthcoming.

\acknowledgments 

We wish to thank for their hospitality the Yukawa Institute and the
organizers of the Post-YKIS2005 mini-workshop on the `Frontiers of
Gravitational Wave Physics' where this work was started.  G.B.C.\
acknowledges support by NSF grant PHY-0555617 and the Z. Smith
Reynolds Foundation.  B.F.W.\ acknowledges support by NSF grant
PHY-0555484. Computations were performed on the Wake Forest University
DEAC Cluster.

\end{document}